# One-dimensional hotspots located between silver nanowire dimers evaluated by surface-enhanced resonance Raman scattering

Tamitake Itoh,[1],* Yuko S. Yamamoto[2,3], Yasutaka Kitahama[4], Jeyadevan Balachandran[5]

[1] Nano-Bioanalysis Research Group, Health Research Institute, National Institute of Advanced Industrial Science and Technology (AIST), Takamatsu, Kagawa 761-0395, Japan

[2] Research Fellow of the Japan Society for the Promotion of Science, Chiyoda, Tokyo 102-8472, Japan

[3] Department of Advanced Materials Sciences, Faculty of Engineering, Kagawa University, Takamatsu, Kagawa 761-0396, Japan

[4] Department of Chemistry, School of Science and Technology, Kwansei Gakuin University, Sanda, Hyogo 669-1337, Japan

[5] Department of Engineering, the University of Shiga Prefecture, Hikone, Shiga 522-8533, Japan

*Corresponding author: tamitake-itou@aist.go.jp

ABSTRACT: Hotspots of surface-enhanced resonance Raman scattering (SERRS) are localized within 1 nm at gaps or crevices of plasmonic nanoparticle (NP) dimers. We demonstrate SERRS hotspots with volumes that are extended in one dimension tens of thousand times compared to standard zero-dimensional hotspots using gaps or crevices of plasmonic nanowire (NW) dimers. According to the polarization measurements, a plasmon resonance in the direction along the dimer width generates SERRS hotspots. SERRS images show oscillating patterns between edges of the hotspot. The SERRS intensity becomes the strongest at the edges, indicating that Fabry–

Perot type resonance of surface plasmons is involved in the Raman enhancement. These optical properties of the SERRS hotspots are quantitatively reproduced by numerical calculations based on the electromagnetic (EM) mechanism. EM coupling energy between dye molecule excitons and plasmons is evaluated using spectral changes in plasmon resonance reflected in a loss of SERRS activity at the hotspots. The coupling energies are consistent with the calculated EM enhancement factors.

Plasmon, which is a collective excitation of conduction elections, can resonate with light and localize the light within tens of nanometers around a metallic nanoparticle (NP).[1] If the two NPs form a dimer, the light is further localized within several nanometers at its gap.[1] Then, the light intensity confined at the gap called "a hotspot" is increased by a factor of approximately $10^5$, substantially enhancing effective light-matter interaction cross-sections.[2,3] Thus, the hotspot effect has been used in surface-enhanced spectroscopy with single molecule sensitivity, including surface-enhanced (resonance) Raman scattering SE(R)RS, surface-enhanced fluorescence (SEF), and their nonlinear counterparts.[4-7] Recently, the hotspots have been attracting considerable attention as tools for examining phenomena related to cavity quantum electrodynamics (QED), e.g., strong coupling between molecular excitons and plasmons, ultrafast SEF, and various other phenomena related to QED.[8-11] To quantitatively evaluate such phenomena, a rigorous investigation of light-matter interactions is needed because several conventional assumptions made in the EM mechanism description, such as a weak coupling approximation, break down in the hotspots.

Unfortunately, volumes of such hotspots are restricted to only several cubic nanometers at the crevices of the plasmonic NP dimers or nanogaps between the plasmonic NP and plasmonic metal surfaces to obtain a single molecule SERRS sensitivity.[12] Thus, such interesting light-matter interactions are quite unstable due to photo-induced thermal and chemical molecular fluctuations.[13] For example, a single molecule SERRS observation is usually interrupted or terminated by desorption and carbonization of molecules at the hotspots.[6,13,14] Thus, it is important to increase the volume of the hotspots to reduce the instability and improve the investigations of the phenomena.

In this work, we investigate the SERRS hotspots formed along the crevices between two parallel silver nanowires (NWs). The length of one-dimensional hotspots is several micrometers to tens of micrometers, which is tens of thousands times longer than that of the hotspots of NP dimers. A polarization dependence of the plasmon resonance parallel to the width of the dimer is the same as that of the SERRS from the hotspots. Oscillating SERRS intensity along the hotspot and the highest SERRS intensity at the edges indicate a contribution of a Fabry–Perot type interference of the surface plasmons propagating through the length of the NW to the SERRS hotspots. These results are quantitatively reproduced by finite-difference time-domain (FDTD) calculations, indicating that the hotspots are generated by the EM enhancement of the plasmon resonance parallel to the dimer width. Spectral changes in the plasmon resonance reflecting a loss of the SERRS activity enabled us to estimate the EM coupling energy between dye molecule excitons and plasmons in the hotspots. The estimated coupling energies are consistent with the calculated EM enhancement factors.

Silver NWs were chemically synthesized through a modified polyol process.[15] To start, 2.45 g of polyvinylpyrrolidone (PVP) was added to 60 ml of ethylene glycol (EG) and heated to

125 °C for 20 min. After the PVP was dissolved in the EG, 0.526 ml of 0.19 M sodium chloride solution was injected into the reactant. Then, 7.65 ml of 0.654 M silver nitrate solution was introduced into the reactant. The reaction was continued for 3 h at 125 °C under continuous stirring at 100 rpm. After the reaction, the suspension was cooled to room temperature naturally. Then, the suspension was centrifuged, and the sediment was collected, washed, and stored in methanol for the morphological analysis and crystal structure analysis using field emission scanning electron microscopy (FE-SEM) and X-ray diffraction (XRD), respectively. The average NW diameter and length were 50 nm and 10 μm, respectively.

The NW suspension was dropped and dried on an indium tin oxide (ITO) glass plate. Then, Rhodamine 6G (R6G) methanol solution ($1.0 \times 10^{-6}$ M) was dropped and dried on the glass plate. The initial dye concentration was substantially higher than that in the standard recipe for single-molecule SERS experiments (~$10^{-10}$ M).[16] Thus, the effective concentration of the dye on the glass surface was reduced by rinsing away most of the dye molecules adsorbed on both the NW and glass surfaces using water and methanol, which was confirmed by blinking signals of the SERRS and fluorescence from both the NWs and glass surface (Movie S1 in Supporting Information). With the blinking signals, we can safely assume that the effective distributions of the dye on the sample surfaces are equivalent to those for near single-molecule SERRS detections.

The experimental setup is illustrated elsewhere.[17] Briefly, a green laser beam (532 nm, 3.5 W/cm$^2$) for a SERRS excitation is focused using the objective lens. White light from a 50 W halogen lamp illuminates the sample through a dark-field condenser (NA 0.8~0.92) to selectively collect an elastic scattering light.[18] The elastic scattering light as well as a SERRS light from an identical NW dimer were collected through an objective lens and sent to a polychromator for the spectroscopy. Detailed information about the experimental setup for the spectral detection of the elastic light scattering is provided in Ref. 18.

The quality of the NWs was examined by the Fabry–Perot type resonance of longitudinal surface plasmon waves.[19] The surface plasmon waves excited by a dark-field illumination at the end of a NW propagate through its length and reflect at the other end. The multiple reflections appear as oscillating patterns in the elastic scattering spectra if the propagating length is reasonably larger than that of the NW.[19] The length is limited by damping due to scattering at surface roughness, domain boundaries, or defects. Figures 1(a1)-(a4) are typical elastic scattering spectra of isolated NW monomers. One can confirm the Fabry–Perot type resonance in the elastic scattering spectra in Figs. 1(a1)-(a4), indicating the NWs can be treated as single-crystalline rather than polycrystalline.

Figures 1(b)-1(d) show the dark-field, SERRS, and SEM images of the NWs for an identical area of the sample ITO glass plate. By comparing these images, one can notice that the NWs with blue, yellow, and pink colors in Fig. 1(b) show the SERRS activity in Fig. 1(c), and these NWs form aggregates shown in Figs. 1(d) and 1(e). The NWs with gray color in Fig. 1(b) do not show any SERRS activity in Fig. 1(c), and these NWs are all monomers, as shown in Fig. 1(d). Note that the plasmon resonance along the length of the NWs is in the near infrared (NIR) region.[20] Thus, the dark-field image shows the elastic light scattering by the plasmon resonance along the NW diameter. Considering that this plasmon resonance is in the violet region,[21] these relationships between the colors of the elastic light scattering, SERRS activity, and aggregation of NWs can be explained as that the aggregate formation by the NWs shifts the plasmon resonance to visible region and as that the SERRS activity exhibited by the aggregates. The results are very encouraging for the realization of one-dimensionally extended hotspots along the gaps of the NW dimers.

To precisely examine the above relationships, an individual NW dimer was investigated using dark-field, SERRS, and SEM images. Figures 2(a)–2(c) show the SEM, dark-field, and SERRS images of the NW dimer. The dimer can be divided into two regions: one is a gap region and other is a monomer region, as shown in Fig. 2(a2) and 2(a3). By superposing the dark-field image on the SEM image as in Fig. 2(d), we determined that only the gap region shows blue color while the monomer region is gray, indicating that the plasmon resonance perpendicular to the NW length red-shifts under the dimerization. By superposing the SERRS image on the SEM image as in Fig. 2(e), we found that only the gap region exhibits the SERRS activity. These results clearly show that a gap region in a NW dimer contains one-dimensionally extended hotspots induced by the red-shifted plasmon resonance caused by the dimerization of the NWs.

To analyze the plasmon resonance that induces one-dimensional hotspots of the SERRS, the polarization dependence of the plasmon resonance and SERRS spectra were measured for a single NW dimer. Figures 3(a) and 3(b) show the elastic light scattering and SERRS spectra perpendicular and parallel to the length of the dimer, respectively. Both the plasmon resonance around 520 nm and SERRS reach the maximum for the polarization perpendicular to the dimer length. Both are minimized for the polarization along the dimer length. In contrast, a broad plasmon resonance around 750 nm, which is similar to the plasmon resonance of the monomer, reaches the maximum for the polarization parallel to the long axis. Figure 3(c) shows the polarization dependence of the plasmon resonance maxima at 520 and 750 nm, respectively. Figure 3(d) shows the polarization dependence of the SERRS. The polarization dependence of the plasmon resonance maxima at 520 nm is the same as that of the SERRS, indicating that the plasmon resonance perpendicular to the length of the dimer induces the EM enhancement generating the SERRS along the dimer crevices.

Dark-field images of the NW aggregates show large color variations (Fig. 1(b)), but the corresponding SERRS images show considerably minor color variations (Fig. 1(c)), despite the fact that even the SERRS spectra of NP dimers exhibit large variations.[22] To understand the origin of the minor variations, we considered the dimer-to-dimer variations in the plasmon resonance and SERRS using the dimer images and their spectra. Figures 4(a1) and 4(a2) show the spectra of the elastic light scattering perpendicular to the dimer length. The elastic light scattering spectra show broad plasmon resonance peaks around 550–600 nm. When a plasmon resonance peak wavelength is longer than approximately 600 nm, a second peak appears around 450–500 nm. Figure 4(c) shows the wavelengths for the first peak vs. the second one. The second peaks appear when the first peak wavelengths become longer than 560 nm, and they red-shift together with the first peaks. Thus, we conclude that the first and second peaks may correspond to a dipole and quadrupole mode of the plasmon resonance, respectively.[23] Figures 4(b1)–4(b2) show the SERRS spectra of the dimers with the SERRS images presented in Figs 4(a1)–4(a2). All SERRS spectra overlap with strong SEF spectra, which are similar to a conventional fluorescence of R6G. The strength of the SEF spectra compared to the SERRS spectra means small SEF quenching rates, pointing to a relatively large distance between the dye molecules and metal surfaces compared with that for NP dimers.[6] Note that the quenching rates are determined by the energy transfer from electronically excited molecules to metal surfaces.[24] The similarity between the SEF and conventional fluorescence implies that the plasmon resonance is too broad to cause spectral modulation in the SERRS and SEF spectra. Indeed, such spectral modulation is clearly observed for SERRS spectra of NP dimers, in which case the plasmon resonance is much narrower than that of NW dimers.[22] We found several interesting physical properties in the SERRS images. The SERRS images in the insets of Figs. 4(b1) and 4(b2) show higher intensity at the edges of the hotspots than in the middle. The SERRS images also show that these hotspots are not uniform, but have some structure along the length of the dimers. Figure 4(d1) shows that the SERRS images clearly exhibit the oscillating patterns, indicating that the Fabry–Perot type interference takes place in the crevice of the NW dimers similar to the NW monomers case. Furthermore, the edges of hotspots maintain a weak SERRS activity even when the incident light polarization is parallel to the dimer length, as shown in Figs. 4(d2)–4(d4).

To analyze the properties of the plasmon resonance and SERRS spectra of the NW dimers presented in Figs. 1–4, we carried out FDTD calculations for the NW dimers. We summarized the experimental findings below.

(1) NWs forming dimers exhibit both the red-shifted plasmon resonance and SERRS activity.

(2) The red-shifted plasmon resonance and SERRS exhibit identical polarization dependence.

(3) Second peaks in elastic scattering spectra appear when the first peak wavelengths exceed 560 nm.

(4) The SERRS intensities become the largest at edges of the hotspots.

(5) The hotspots show the oscillating patterns along the dimer length.

(6) The SERRS intensities at the edges of hotspots do not disappear even when the incident light polarization is parallel to the dimer length.

(7) The SERRS intensities and spectra exhibit only minor dimer-to-dimer variations.

Figure 5(a) shows the structure of an NW dimer used for the FDTD calculations to understand the seven above-mentioned properties. The validation of the calculations is described elsewhere.[3]

The length and diameter of each NW in the dimer are 6.0 μm and 100 nm, respectively. The NWs are contacting each other. The length of the crevice between the NWs is 5.0 μm. Figure 5(b) shows spectra of the elastic scattering perpendicular and parallel to the dimer length, and the inset shows those spectra for the monomer. When the polarization is perpendicular to the dimer length, both dipole and quadrupole plasmon resonance contributions appear around 620 and 460 nm, respectively. They disappear when the polarization is along the dimer length. These plasmon resonances do not appear in the monomer. These spectral properties support well the experimental results in Fig. 3, indicating that the calculations correctly reproduce the EM fields around the dimers. Figures 5(c)–5(e) show the EM field distributions in X, Y, and Z planes at the excitation wavelength of 532 nm, respectively. Figures 5(c2) and 5(c3) show that an enhanced EM field appears at the crevice of the dimer when the polarization is oriented along the dimer width (Z axis), and the EM field disappears when the polarization is parallel to the dimer length (X axis). Figures 5(d2)–5(d3) and 5(e2)–5(e3) show Fabry–Perot type oscillating patterns of the EM fields. The EM field intensity reaches the maximum at the edges of the crevices for the polarization oriented along the dimer width. Taking into account the incident light polarization and generation of one-dimensional hotspots, the surface plasmon involved in the oscillating patterns in the calculations should give rise to transverse anti-symmetric surface plasmon waves propagating along the dimer length.[25] Figures 5(d3) and 5(e3) clearly show the largest enhancement of the EM fields at the ends of the crevices even when the polarization is parallel to the dimer length. The calculation results can be explained as follows. A non-radiative longitudinal surface plasmon wave gives rise to the radiative local plasmon by contacting the edges of the crevices. The local plasmon generates the enhanced EM field and supports the contribution of the Fabry–Perot type resonance to the SERRS.[26] These calculation results well reproduce the experimental properties (1)-(6), indicating that the calculations correctly reproduce the EM fields generating the SERRS hotspots.

To evaluate the spectral variations in the plasmon resonance and SERRS of the NW dimers, we carried out an FDTD calculation of the NW dimers for different NW diameters. Figure 6(a) shows the spectra of the elastic light scattering parallel to the dimer width for the diameters of both NWs in the range of 50–110 nm. The length of the NWs is set to 6.0 µm. As increase in the diameter causes a red shift in the dipole plasmon resonance and generation of quadrupole plasmon resonance, which appears at wavelengths shorter than 450 nm. These variations in the elastic scattering spectra are consistent with the experimental results in Fig. 4(c). We obtained the enhancement factor $M$ of the EM field amplitude at the crevices of these NW dimers. Figures 6(b) and 6(c) show the spectra of $M$ at the centers and ends of the dimer crevices, respectively. The values of $M$ at 532 nm are around 70 and 150 at the centers and ends, respectively. The values of $M$ exhibit only minor dimer-to-dimer variations around 532 nm even for different NW diameters. These minor variations may be due to the broad line widths of the dipole plasmon resonance, and they are consistent with the experimental property (7) showing minor variations in the SERRS intensities (Fig. 1(c)).

We investigated the EM coupling between the plasmon and molecular excitons by analyzing the spectral changes in the plasmon resonance of the NW dimers. In the case of NP dimers, this was accomplished by comparing the experimental elastic scattering spectra before and after a loss of the SERRS activity.[27] Figures 7(a1)–7(a3) show changes in the elastic scattering spectra of a dimer before a loss (Fig. 7(a1)), after a loss (Fig. 7(a2)), and after a recovery (Fig. 7(a3)) of the SERRS activity. The SERRS activity was quenched by rinsing away dye molecules from the NW dimers with methanol and recovered by putting the dye solution on them. A loss or recovery of the SERRS activity is manifested in blue or red spectral shifts of the plasmon resonance. The reproducibility of the spectral shifts indicates that they are not caused by the structural changes in the NW dimers. Figure 7(b) shows the relationships between the peak energies and widths of the plasmon resonance. One can notice that the loss of the SERRS activity results in the blue shifts and narrowing of the peaks. We concluded that these spectral changes are induced by the decreasing EM coupling energy between the plasmonic and molecular electronic resonances. Indeed, a loss of the SERRS activity is identical to a disappearance of the EM coupling energy.[17] We applied a cavity QED model to quantitatively evaluate the decrease.[28] In this model, the elastic light scattering spectra are calculated using a model of a classical coupled-oscillator composed of the plasmonic and molecular oscillators representing the electronic resonance of a two-level system.[28] The equations of motion for the two oscillators are

$$\frac{\partial^2 x_p(t)}{\partial t^2} + \gamma_p \frac{\partial x_p(t)}{\partial t} + \omega_p^2 x_p(t) + g \frac{\partial x_m(t)}{\partial t} = F_p(t) \tag{1}$$

$$\frac{\partial^2 x_m(t)}{\partial t^2} + \gamma_m \frac{\partial x_m(t)}{\partial t} + \omega_m^2 x_m(t) - g \frac{\partial x_p(t)}{\partial t} = F_m(t), \tag{2}$$

where $x_p$ and $x_m$ are coordinates of the plasmonic and molecular electronic oscillations, respectively; $\gamma_p$ and $\gamma_m$ are line-widths of the plasmonic and molecular electronic resonances, respectively; $\omega_p$ and $\omega_m$ are resonance frequencies of the plasmon and molecular electronic transitions, respectively; $g$ is a coupling rate, and $F_p$ and $F_m$ are respective driving forces representing the incident light. We set $F_m(t) = 0$, because the coupled-oscillator is almost entirely driven by the excitation light energy obtained by the plasmonic oscillator, $F_m(t) \ll F_p(t)$. By assuming $F_p(t) = F_p e^{-i\omega t}$, where $\omega$ is the incident light frequency, $x_p(t)$ and $x_m(t)$ can be derived from Eqs. (1)–(2). In the quasi-static limit, the scattering cross-section $C_{sca} = (8\pi/3k)|\alpha|^2$, where $k$

= ωn/c is the wave vector of light and α = $F_p x_p$ is the polarizability of the dimer. By substituting $x_p(t)$ into α, we can obtain $C_{sca}$ as

$$C_{sca}(\omega) = \frac{8\pi}{3}k|\alpha|^2 \propto \omega^4 \left| \frac{\omega_m^2 - \omega^2 - i\gamma_m\omega}{(\omega^2 - \omega_p^2 + i\gamma_p\omega)(\omega^2 - \omega_m^2 + i\gamma_m\omega) - \omega^2 g^2} \right|^2. \quad (3)$$

We analyzed the experimental results using Eq. (3). The upper panel of Fig. 7(c) shows a typical experimental result for the spectral change of the elastic light scattering caused by the loss of the SERRS activity. The spectral peak exhibits a slight blue shift, and its width becomes slightly narrower (black line in Fig. 7(c1)). The lower panel of Fig. 7(c) shows calculated changes in the elastic light scattering spectra caused by a decrease in the coupling energy from 100 to 0 meV. The experimental blue shift and narrowing caused by the loss of the SERRS activity are well reproduced by the decrease in the coupling energy. We examined this relationship using Eq. (3). Figure 7(d) shows the values of these peak energies and line widths of the elastic light scattering spectra for the coupling energies $\hbar g$ changing from 150 to 0 meV. The variations in blue shifts and broadening are within the experimentally observed range (Fig. 7(b)), indicating that the dimer-to-dimer fluctuations of both plasmon resonance energy and coupling energy are the origin of the variations in the experimental elastic light scattering spectra.

Finally, we estimated the EM enhancement factors using the value of the coupling energy. In the approximation $\omega_0 = \omega_p = \omega_m$ and $\gamma_0 = \gamma_p = \gamma_m$, the value of the spectral splitting is estimated using the oscillator strength (OS) $f = 2m\omega_0 d^2/(e^2\hbar)$ as [29]

$$g = \left( \frac{1}{4\pi\varepsilon_r\varepsilon_0} \frac{\pi e^2 N f}{mV} \right)^{1/2}, \quad (4)$$

where $\varepsilon_r$ and $\varepsilon_0$ are the relative permittivity 1.77 and vacuum permittivity ~8.854 × $10^{-12}$ F/m, respectively; $N$ is the number of R6G molecules at a hotspot of a dimer crevice; $f$ of a R6G molecule is 0.69;[30] e is the electron charge $\sim 1.602 \times 10^{-19}$ C; $m$ is free electron mass $\sim 9.109 \times 10^{-31}$ kg; and $V$ ($\sim LA^2$) is an effective mode volume of the plasmonic cavity at a dimer crevice, where $L$ is the length of the crevice and $A$ is the length of the mode cross-section.[31] To achieve the typical blue shifts $\sim 100$ meV, $V/N$ should be $\sim 8.0 \times 10^{-6}$ $(\lambda/n)^3$, where $\lambda = 600$ nm and $n = 1.33$. To estimate $LA^2$, we set the NW dimer diameter to 40 nm and its crevice length to $L = 5$ μm. The area per one R6G molecule is estimated as $1.6 \times 10^2$ nm$^2$ from the sample preparation conditions (the R6G concentration of $1.0 \times 10^{-6}$ M, a droplet amount of 10 μ$l$, and an area of a cover glass $2 \times 5$ cm$^2$) by assuming the uniform dispersion of R6G molecules on the substrate. During a drying process, molecules may be concentrated into a crevice. By assuming a half of molecules on a NW dimer are concentrated into the crevice, the number of molecules $N$ at the crevice is around 5000. Thus, from $LA^2/N \sim 8.0 \times 10^{-6}$ $(\lambda/n)^3$, the value of $A$ is estimated as 26 nm. The value of 26 nm is much larger than the minimum value $\sim 1.7 \times 10^{-8/3}$ $(\lambda/n)$,[32] indicating that the observed spectral changes can be generated by the loss of the dye molecules at the crevice of the dimer. Furthermore, the inverse of the value $8.0 \times 10^{-6}$, $(8.0 \times 10^{-6})^{-1} \sim 10^5$, which roughly corresponds to an enhancement factor of the EM field intensity $|M|^2$ in Figs. 6(b) and 6(c), is approximately consistent with the estimated enhancement factor of the EM field intensity obtained from the FDTD calculations $\sim 2 \times 10^4$ (Fig. 6).

    In this letter, we examined one-dimensionally extended SERRS hotspots generated using gaps of silver NW dimers. The comparison between experiments and FDTD calculations quantitatively determined that a plasmon resonance in the direction along the dimer width generates the SERRS hotspots, and that oscillating patterns in SERRS images are induced by a Fabry–Perot type interference of the propagating surface plasmons. The EM coupling energies for the coupling between dye molecule excitons and plasmons were determined to be around 100

meV. We obtained them from the spectral changes in the plasmon resonance caused by a loss of the SERRS activity. The experimental values of EM enhancement factors are consistent with those from the FDTD calculations.

This study was supported in part by WAKATE B (No. 26810013) and KIBAN B (No. 26286066) grants from the Ministry of Education, Culture, Sports, Science, and Technology of Japan, and by an A-STEP Grant (No. AS2525017J) from the Japan Science and Technology Agency (JST).

**REFERENCES**

[1] H. Xu, J. Aizpurua, M. Käll, P. Apell, Phys. Rev. E **62**, 4318 (2000).

[2] P. Johansson, H. Xu, M. Käll, Phys. Rev. B **72**, 035427 (2005).

[3] K. Yoshida, T. Itoh, H. Tamaru, V. Biju, M. Ishikawa, and Y. Ozaki, Phys. Rev. B **81,** 115406 (2010).

[4] K. Kneipp, Y. Wang, H. Kneipp, L. Perelman, I. Itzkan, R. R. Dasari, M. Feld, Phys. Rev. Lett. **78**, 1667 (1997); S. Nie, S. Emory, Science **275**, 1102 (1997).

[5] E. C. Le Ru, P. G. Etchegoin, J. Grand, N. Fe´lidj, J. Aubard, and G. Le´vi, J. Phys. Chem. C **111**, 16076 (2007).

[6] T. Itoh, M. Iga, H. Tamaru, K. Yoshida, V. Biju, and M. Ishikawa, J. Chem. Phys. **136**, 024703 (2012).

[7] T. Itoh, H. Yoshikawa, K. Yoshida, V. Biju, M. Ishikawa, J. Chem. Phys. **130**, 214706 (2009).

[8] A. E. Schlather, N. Large, A. S. Urban, P. Nordlander, N. J. Halas, Nano Lett. **13,** 3281 (2013).

[9] F. Nagasawa, M. Takase, K. Murakoshi. J. Phys. Chem. Lett. **5**, 14 (2014).

[10] T. Itoh, Y. S. Yamamoto, H. Tamaru, V. Biju, N. Murase, Y. Ozaki, Phys. Rev. B **87,** 235408 (2013).

[11] Y. S. Yamamoto, Y. Ozaki, and T. Itoh, J. Photochem. Photobio. C **21**, 81 (2014).

[12] K. Imura, H. Okamoto, M. Hossain, M. Kitajima, Nano Lett. **6**, 2173 (2006).

[13] M. Moskovits, L. L. Tay, J. Yang, and T. Haslett, Top. Appl. Phys. **82**, 215 (2002); T. Itoh, Y. S Yamamoto, Analyst **141**, 5000 (2016).

[14] T. Itoh, Y. S Yamamoto, V. Biju, H. Tamaru, S. Wakida, AIP Adv. **5**, 127113 (2015).

[15] B. Li, S. Ye, I. E. Stewart, S. Alvarez, and B. J. Wiley, Nano Lett. **15**, 6722 (2015).

[16] A. B. Zrimsek, N. L. Wong, and R. P. Van Duyne, J. Phys. Chem. C **120**, 5133−5142 (2016).

[17] T. Itoh, Y. S Yamamoto, H. Tamaru, V. Biju, S. Wakida, and Y. Ozaki, Phys. Rev. B **89**, 195436 (2014).

[18] T Itoh, Y. S Yamamoto, T Suzuki, Y Kitahama, Y Ozaki, Appl. Phys. Lett. **108**, 021604 (2016).

[19] H. Ditlbacher, A. Hohenau, D. Wagner, U. Kreibig, M. Rogers, F. Hofer, F. R. Aussenegg, and J. R. Krenn, Phys. Rev. Lett. **95**, 257403 (2005).

[20] F. Neubrech, T. Kolb, R. Lovrincic, G. Fahsold, A. Puccia, J. Aizpurua, T. W. Cornelius, M. E. Toimil-Molares, and R. Neumann, S. Karim, Appl. Phys. Lett. **89**, 253104 (2006).

[21] J. P. Kottmann, O. J. F. Martin, D. R. Smith and S. Schultz, Phys. Rev. B **64**, 235402 (2001).

[22] K. Yoshida, T. Itoh, V. Biju, M. Ishikawa, Y. Ozaki, Phys. Rev. B **79**, 085419 (2009); Y. S Yamamoto, T. Itoh, J. Raman Spectrosc. **47**, 78-88 (2016).

[23] M. Fu, L. Qian, H. Long, K. Wang, P. Lu, Y. P. Rakovich, F. Hetsch, A. S. Sushad and A. L. Rogachd, Nanoscale **6**, 9192–9197 (2014).

[24] C. M. Galloway, P. G. Etchegoin, E. C. Le Ru, Phys. Rev. Lett. **103**, 063003 (2009).

[25] L. J. E. Anderson, C. M. Payne, Y. Zhen, P. Nordlander, and J. H. Hafner, Nano Lett. **11**, 5034–5037 (2011).

[26] T. Shegai, V. D. Miljkovic, K. Bao, H. Xu, P. Nordlander, P. Johansson, and M. Kall, Nano Lett. **11**, 706–711 (2011).

[27] T. Itoh, K. Hashimoto, A. Ikehata, and Y. Ozaki, Appl. Phys. Lett. **83**, 5557 (2003).

[28] X. Wu, S. K. Gray, and M. Pelton, Opt. Express, **18**, 23633 (2010).

[29] L. C. Andreani, G. Panzarini, J. M. Gerard, Phys. Rev. B **60,** 13276 (1999).

[30] S. Delysse, J. M. Nunzi and C. Scala-Valero, Appl. Phys. B **66,** 439 (1998).

[31] D. E. Chang, A. S. Sørensen, P. R. Hemmer, and M. D. Lukin, Phys. Rev. B, **76**, 035420 (2007).

[32] K. J. Savage, M. M. Hawkeye, R. Esteban, A. G. Borisov, J. Aizpurua, J. J. Baumberg, Nature, **491**, 574 (2012).

**Figure captions**

Fig. 1 (a1)–(a4) Elastic light scattering spectra of isolated NWs. Insets: dark-field images (left panels) and enhanced dark-field images (right panels), respectively. Spectra correspond to white circles in dark-field images of the insets. Images of (b) the elastic light scattering, (c) SERRS, and (d) SEM of the NWs obtained from the same area on the sample glass plate covered with an ITO film. (e) Enlarged SEM images of the NWs corresponding to white boxes (1)–(3) in (b)–(d). Scale bars in (a)–(d) and (e) are 10 µm and 1.0 µm, respectively.

FIG. 2 (a1) SEM image of the NW dimer. Insets: (a2)–(a3) enlarged SEM images of the edges of the dimer crevice. (b)–(c) Dark field and SERRS images of the dimer. (d) Superposition of the SEM and the dark-field image. (e) Superposition of the SEM and SERRS images. Scale bars in (a)–(c) are 5 µm.

FIG. 3 (a) Spectrum of the elastic light scattering for the NW dimer and (inset) monomer perpendicular (red line) and parallel (black line) to the dimer length. (b) Spectrum of the SERRS intensity perpendicular (red line) and parallel (black line) to the dimer length. (c) Polarization dependence (circles and line) of the plasmon resonance maxima at 520 nm (red) and 720 nm (black). (d) Polarization dependence of the SERRS intensity at 560 nm (red circles and red line).

FIG. 4 (a1)–(a2) Spectra of the elastic light scattering efficiency for the direction perpendicular to the NW dimer length and (insets) corresponding dark-field images. (b1)–(b2) Spectra of the SERRS intensity for the direction perpendicular to the NW dimer length and (insets) corresponding images. Red and blue spectra correspond to red and blue circles in insets, respectively. (c) First peak wavelengths vs. the second peak wavelengths in the elastic light scattering spectra of the NW dimers; (○) and (×) denote the presence and absence of the second peak, respectively. (d1) SERRS image of the NW dimer showing spatial oscillatory patterns along the NW dimer length. (d2)–(d3) SERRS images perpendicular and parallel to the NW dimer length, respectively. (d4) The enhanced SERRS image of (d3). Scale bars in (a) – (d) are 5.0 µm.

FIG. 5 (a) Image of a NW dimer and direction of the incident light (yellow arrow) with the coordinate system used for the FDTD calculations. Red and black arrows indicate the direction of the light polarization. (b) Calculated spectra of the elastic light scattering for the NW dimer and (inset) monomer perpendicular (red line) and parallel (black line) to the dimer (monomer) length (X axis). (c1)–(e1) Images of the dimer cross-section in X, Y, and Z planes, respectively. Red and black arrows indicate the direction of the polarization. (c2)–(e2) The EM field distributions at 532 nm for the incident light polarization along the dimer width (Z axis). (c3)–(e3) The EM field distributions at 532 nm for the incident light polarization along X.

FIG. 6 (a) The NW dimer scattering cross-section spectra for different NW diameters for the elastic light scattering in the direction perpendicular to the dimer length: (○), (◊), (▽), (△), and (□) correspond to the diameters of 110, 100, 90, 70, and 50 nm, respectively. Inset: The peak wavelength of the dipolar plasmon resonance vs. the NW diameter. (b) The spectra of the maximum EM field amplitude in the X plane and (c) Z plane for the incident light polarization perpendicular to the dimer length for the NW diameters in (a). Insets: Respective maximum EM field amplitudes at 532 nm vs. the NW diameter.

FIG. 7 Elastic light scattering spectrum of the NW dimer (a1) showing the SERRS activity (a2) after a loss of the SERRS activity, (a3) after recovery of the SERRS activity. (b) Line width vs. peak energy for the elastic light scattering spectra of the NW dimers (red) showing the SERRS activity and (black) after a loss of the SERRS activity. (c1) Experimental elastic light scattering spectra of the NW dimer (red line) showing the SERRS activity and (black line) after a loss of the SERRS activity. (c2) Elastic light scattering spectrum calculated using Eq. (3) with the coupling energy of 100 meV (red line) and 0 meV (black line). (d) Line width vs. peak energy for the elastic light scattering spectra calculated using Eq. (3) with the coupling energy changing from 150 to 0 meV. The values of the plasmon resonance energies for 0 meV coupling energy are (▼) 2.14, (◊) 2.21, and (○) 2.29 eV for the width of 580 meV, and (□) 2.21, (△) 2.29 eV for the width of 490 meV.

Fig. 1

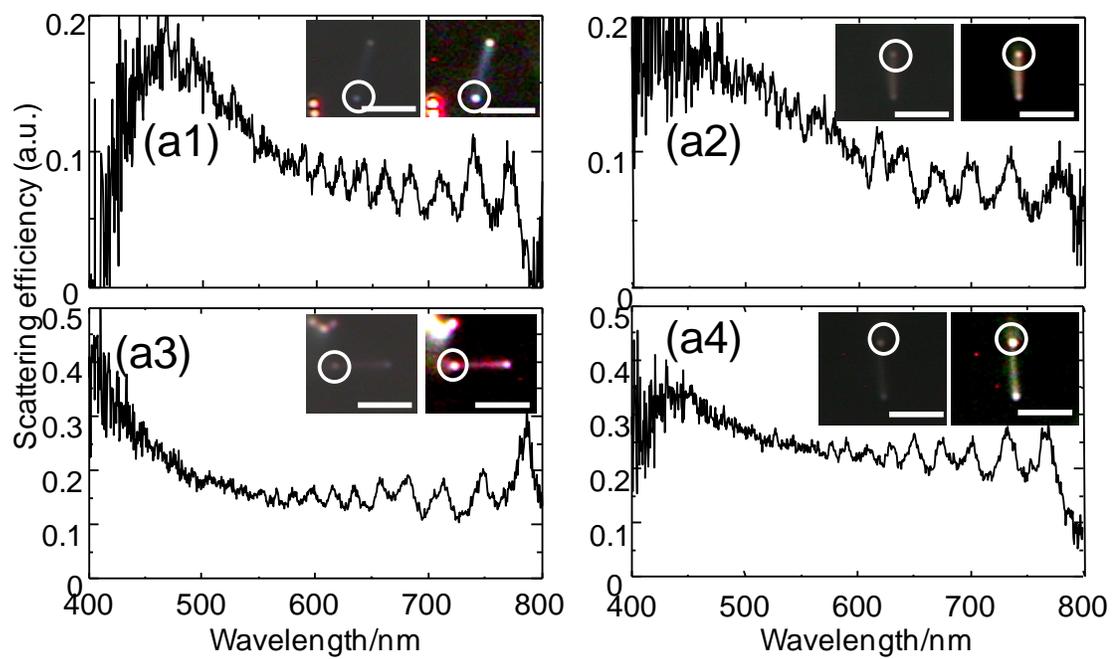
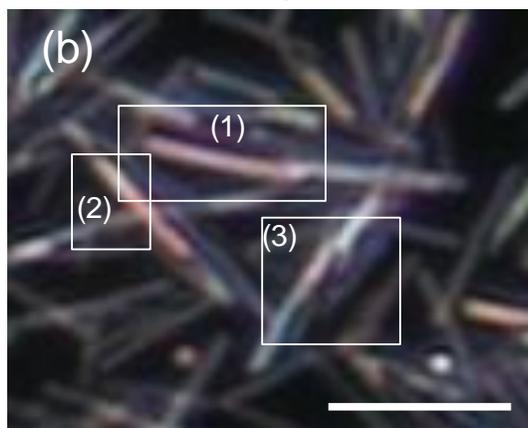
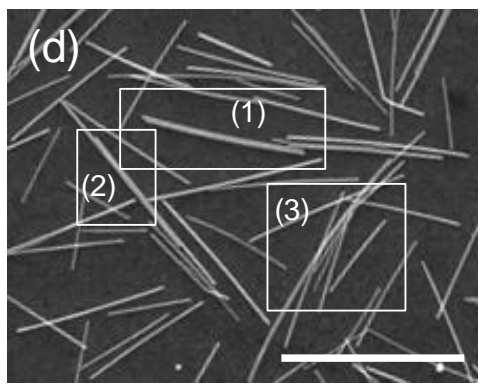
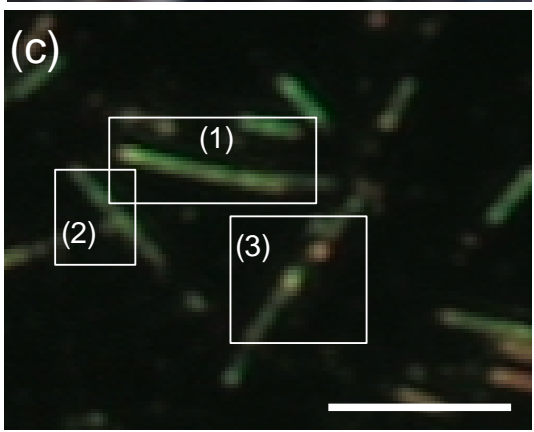
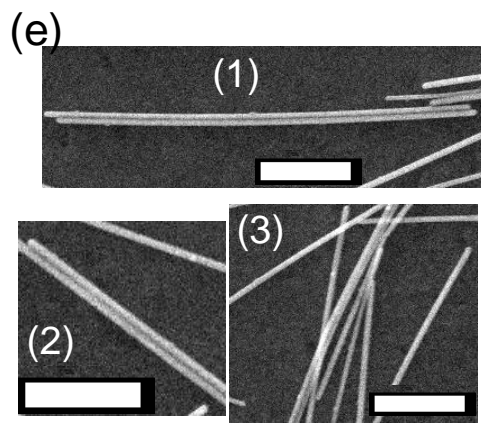

Fig. 2

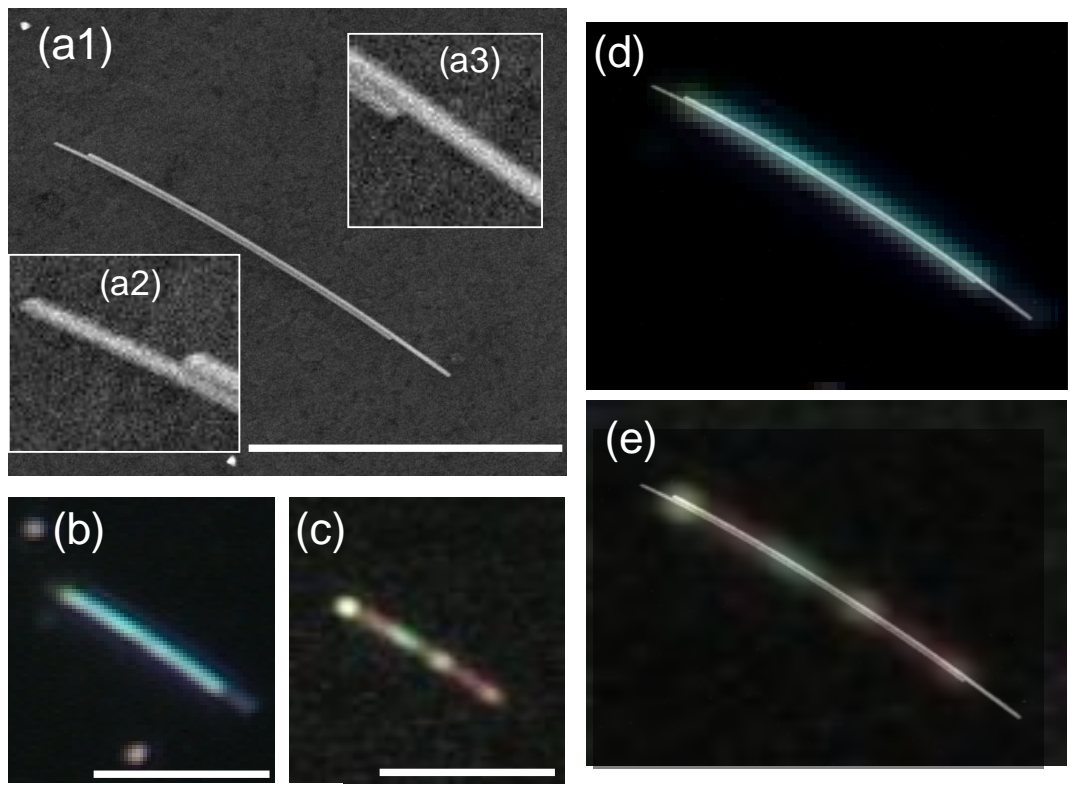

Fig. 3

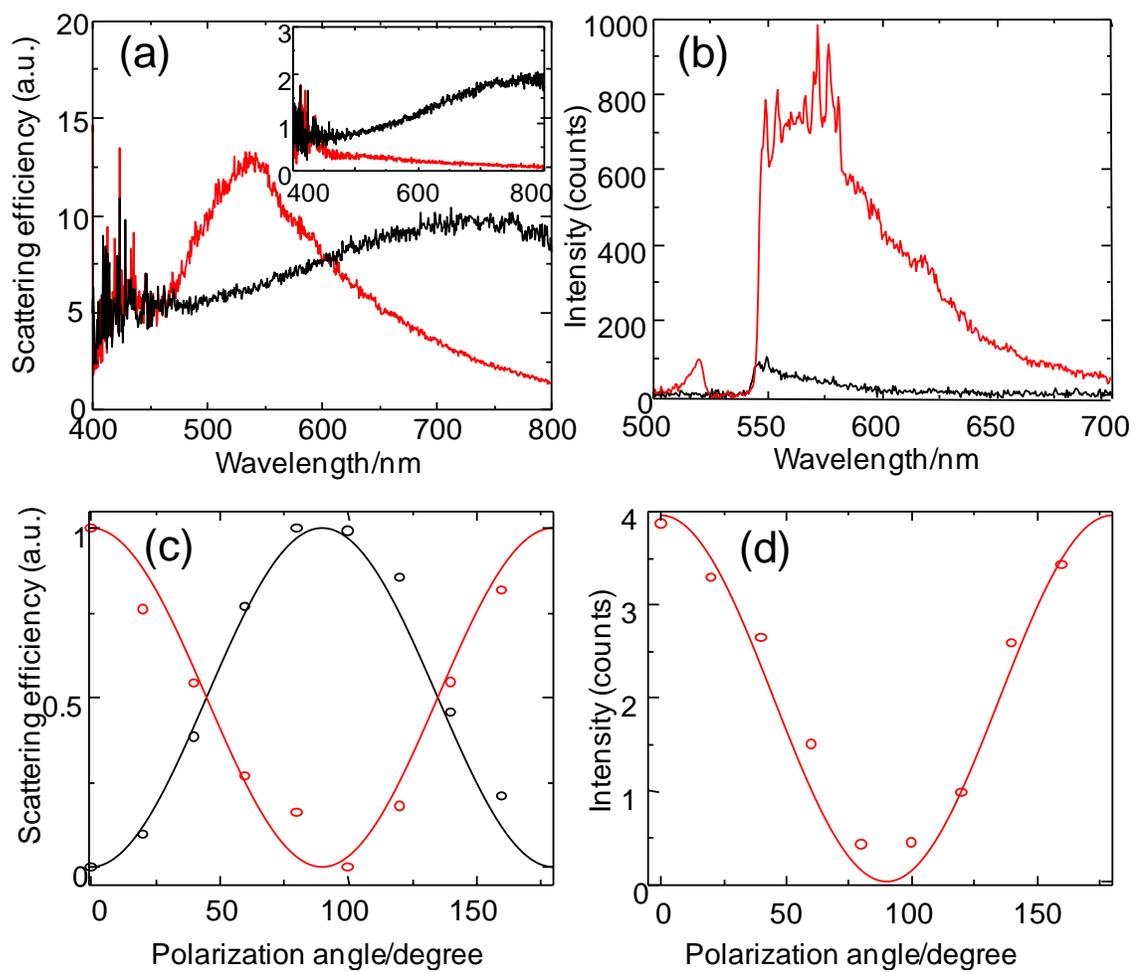

Fig. 4

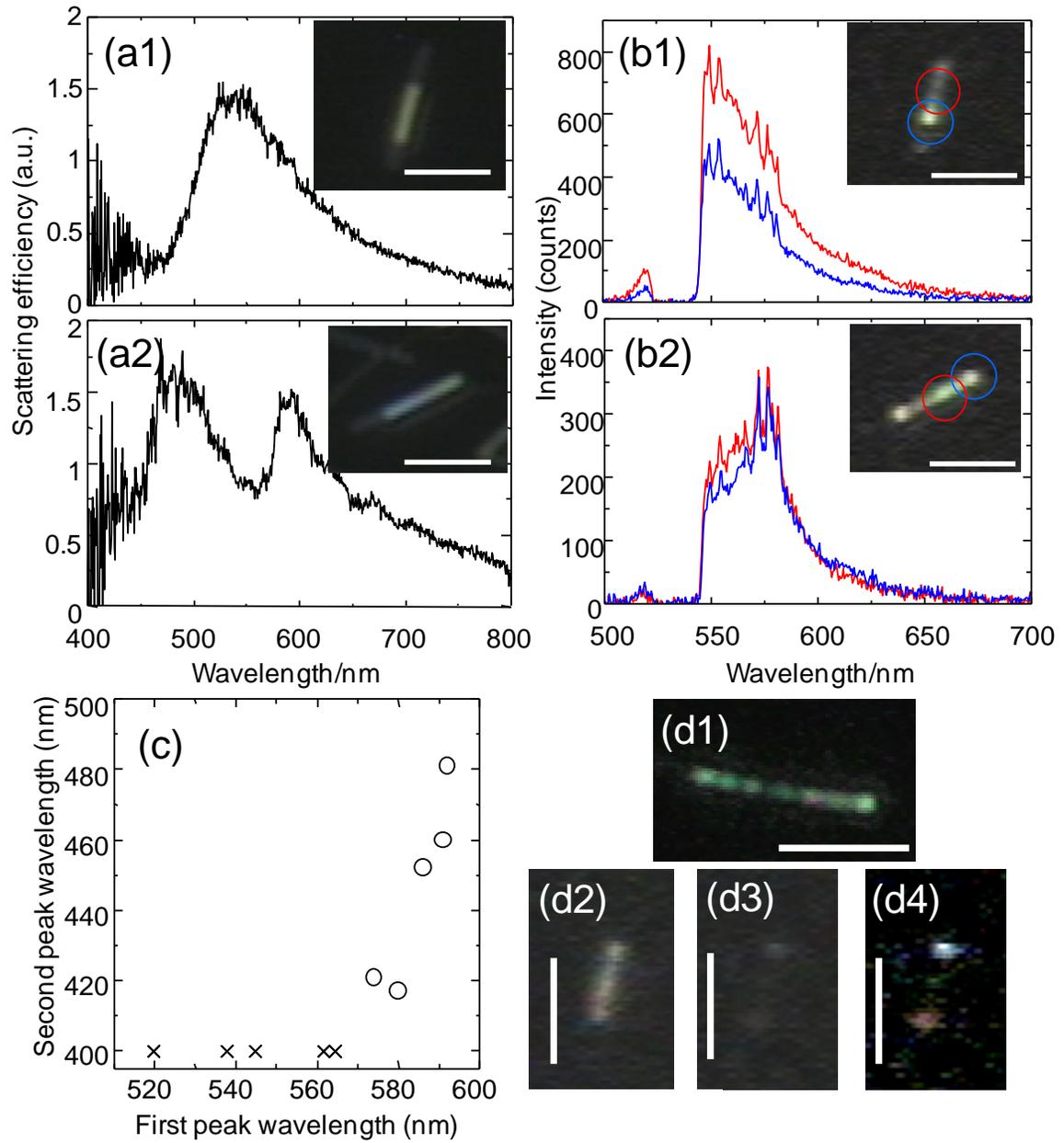

Fig. 5

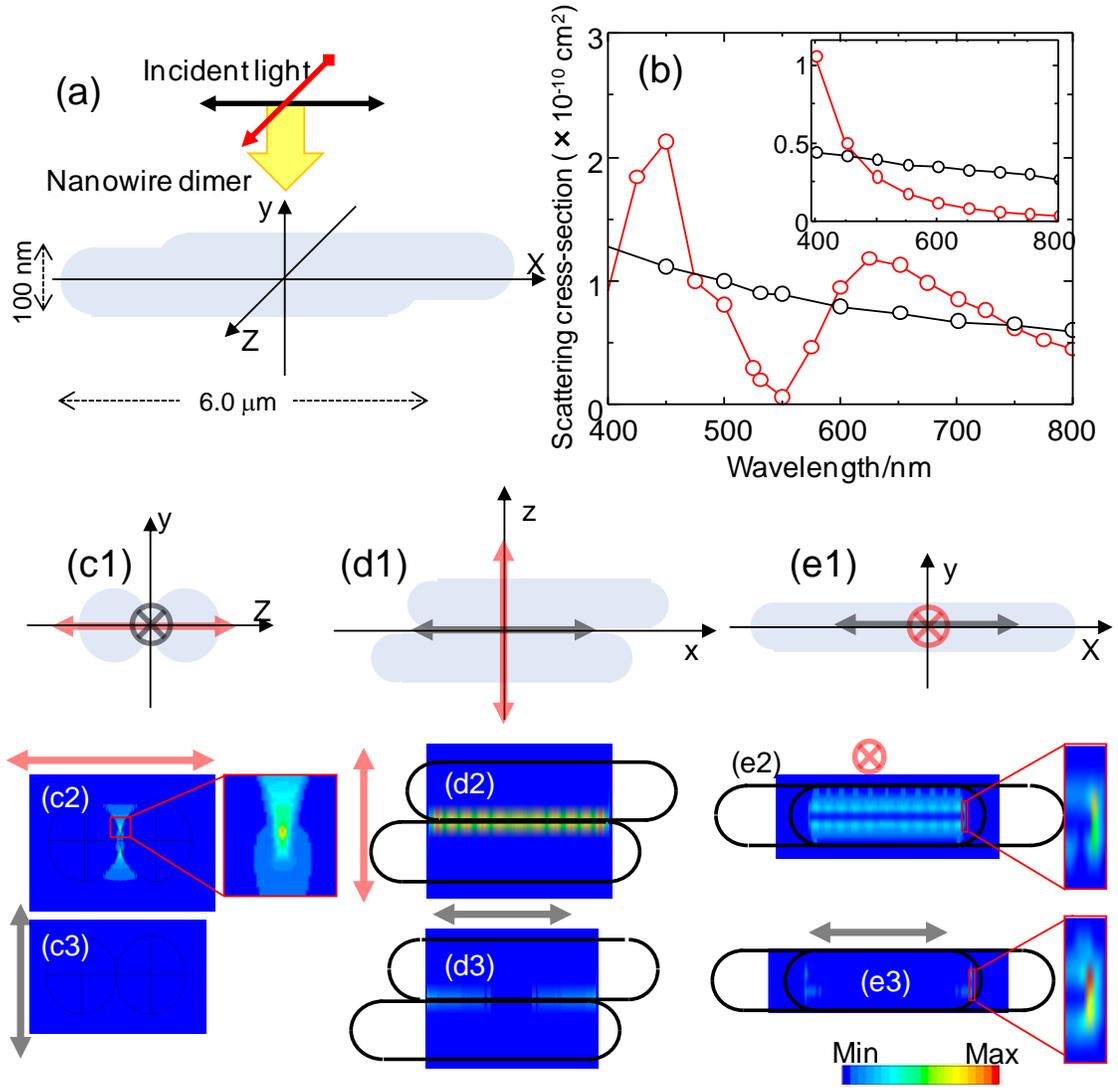

Fig. 6

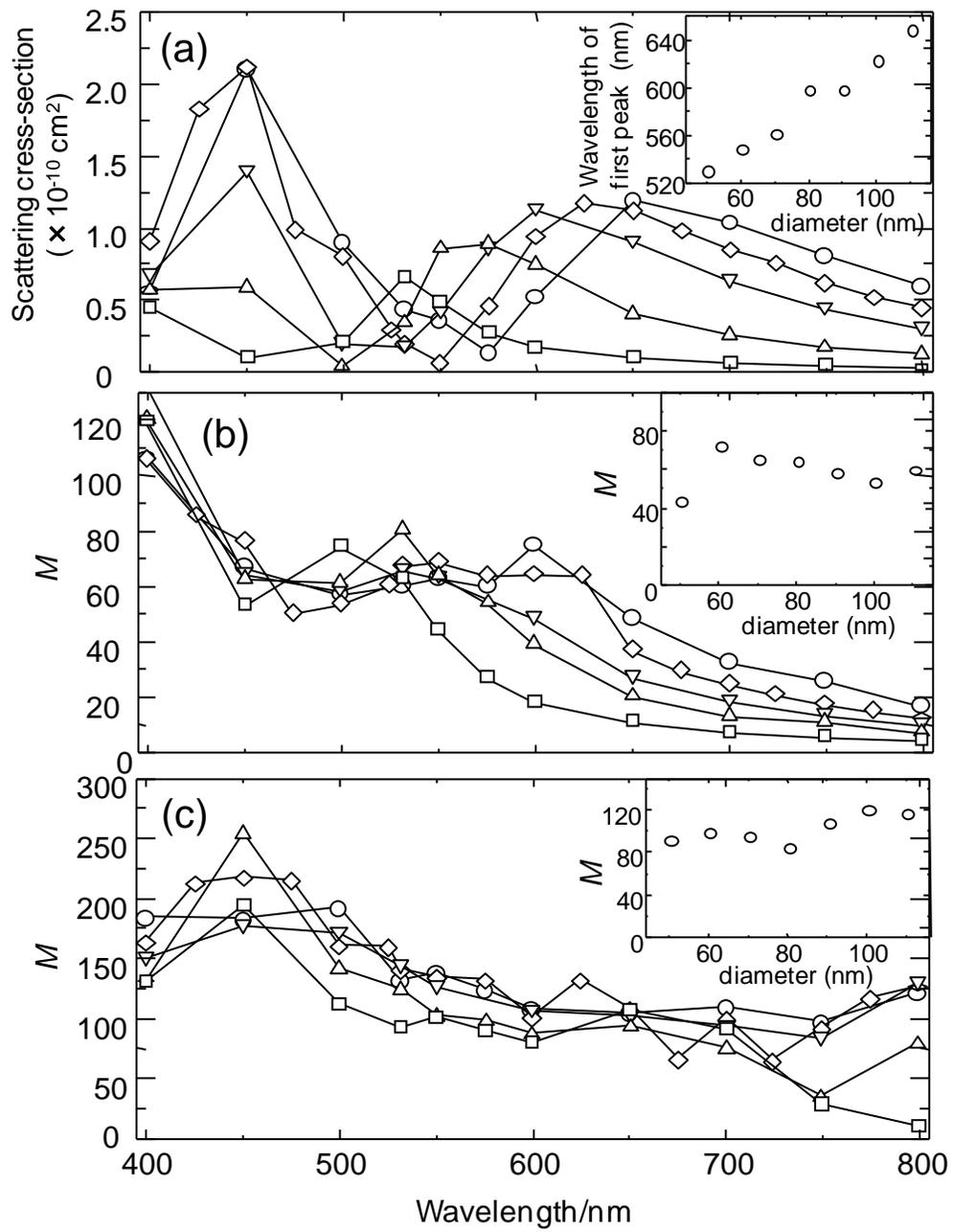

Fig. 7

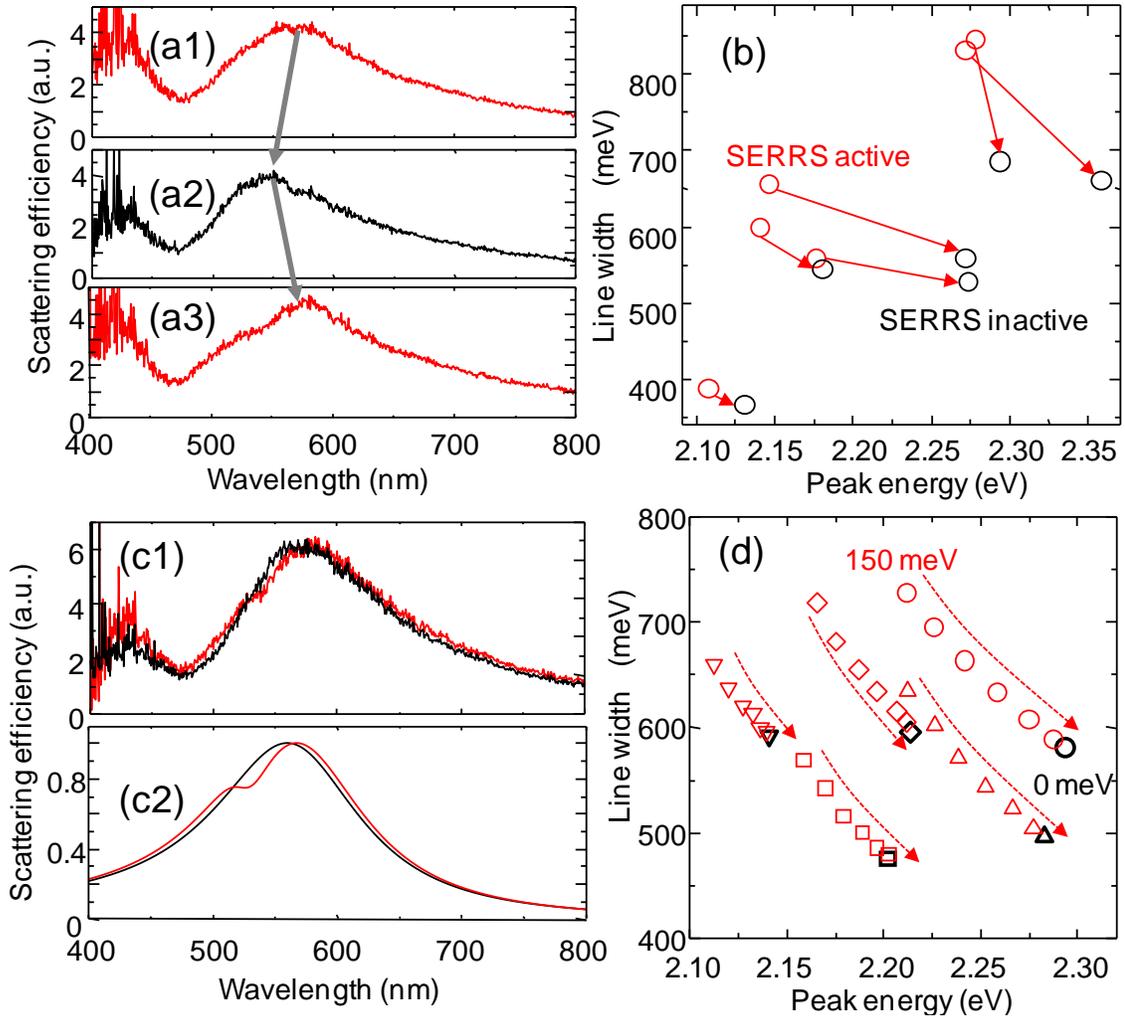